# Optical Properties of Dilute CuAl$_{1-x}$Fe$_x$O$_2$ Delafossite Alloys

M. Aziziha[1], S. Akbarshahi[1], S. Ghosh[2], P. Pramanik[2], A.H. Romero[1], S. Thota[2], S. Pittala[3], M.S. Seehra[1], and M.B. Johnson[1]

[1]Department of Physics & Astronomy, West Virginia University, Morgantown, WV 26506, USA
[2]Department of Physics, Indian Institute of Technology, Guwahati, Assam 781039, India
[3]Department of Physics, Indian Institute of Science, Bangalore 560012, India

**Abstract:** For powder samples of CuAl$_{1-x}$Fe$_x$O$_2$ ($x$ = 0, 0.01, 0.05, and 0.1), measured optical properties are compared with model simulations and phonon spectra are compared with simulations based on weighted dynamical matrix approach.

Dilute alloys of complex oxides hold the promise of exotic properties, such as superconductivity or spin and quantum behavior; as well as practical applications, such as transparent conductive oxides and photo-catalysts. Modern investigations into such material systems involve density functional theory (DFT) calculations done hand-in-hand with crystal synthesis/growth. However, detailed DFT calculations can be computationally expensive, especially for dilute alloys of complex crystals and their phonon properties. Here we report the structural and optical properties of dilute magnetic CuAl$_{1-x}$Fe$_x$O$_2$ DFT modeling, synthesis, and measured experimentally. The details of the elemental and chemical characterization and magnetic properties of alloys are reported recently [1,2].

Powder samples of CuAl$_{1-x}$Fe$_x$O$_2$ ($x$ = 0, 0.01, 0.05, and 0.1) were prepared by the solid-state reaction method using Cu$_2$O, Al$_2$O$_3$, and α-Fe$_2$O$_3$ as precursors in appropriate stoichiometric ratios with a heat treatment in static air at 1,100 °C for 144 h. Raman and UV-Vis-IR measurements used powder reflection configurations, and FTIR used KBr pellets [3].

Lattice parameters are calculated for CuAl$_{1-x}$Fe$_x$O$_2$ in ground state energy using the Vienna Ab Initio Simulation Package (VASP 5.4.4.) Due to the strong correlation effects of Fe $d$-electrons, we used the PBEsol+$U$ approach [3], including Fe spin (see [3] for details). Optimized $U$ is found by comparing to experimental values for CuFeO$_2$. Both atomic masses and force constants play a significant role in phonon modes. Hence, we developed a new effective-medium approach by constructing the dynamical matrix using the weighted averages of the mass and force constants of the parent compounds CuAlO$_2$ and CuFeO$_2$ [3].

X-ray diffraction, including Rietveld refinement, was used to characterize the powders. Figure 1.a shows the variation of the lattice parameters $a$ and $c$ with the change in Fe-doping. The solid lines show the Vegard's law fit. Using the optimized DFT (supercells shown in Figs. 1.b-c) we achieve good agreement with the experimental data and Vegard's law fit.

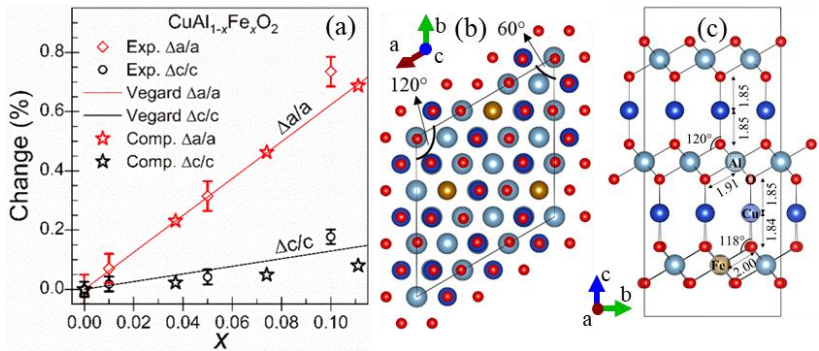

Fig. 1. (a) Experimental data (with error bars) for change of $a$ and $c$ vs. $x$ and computational lattice parameters for CuAl$_{1-x}$Fe$_x$O$_2$ (stars). Lines show the fit of the experimental data to Vegard's law (b) and (c) show the minimum energy configuration for $x$ = 0.11 along, respectively, $a$ and $c$-axes.

Optical properties are investigated by using this spin-polarized PBEsol+$U$ to calculate the electronic band structure for CuAl$_{1-x}$Fe$_x$O$_2$ ($x$ = 0, 0.04, 0.08, and 0.12). (The fractions used for modeling are set by the size of the supercell chosen and are close enough to $x$ = 0.05 and 0.10, the fractions for two of our synthesized powders.) In Figure 2.a and b we show the modeled band structure for $x$ = 0 and $x$ = 0.04 (for both spins), and in Fig. 2.c we show the absorption spectra for as determined from reflectivity measurements for $x$ = 0 and $x$ = 0.05, after nonlinear background subtraction. Absorption spectra for synthesized powders include peaks from various defects (see [4,5]). Disregarding these peaks, the notable difference between $x$ = 0 and $x$ = 0.04 is the appearance of peak **B** at 1.4 eV. (A similar peak is also experimentally observed for $x$ = 0.01.) This is interpreted as resulting from the appearance of the Fe-related states (shown in blue in Fig. 2.b) appearing below the conduction band in the spin-down band diagrams (outlined in the dashed cyan box). The model and experiment show a slight redshift of this Fe-related feature with increasing $x$. Peak **F**, observed in both spectra, is from the direct bandgap shown in the band diagrams for both $x = 0$ and 0.04 (solid green arrows).

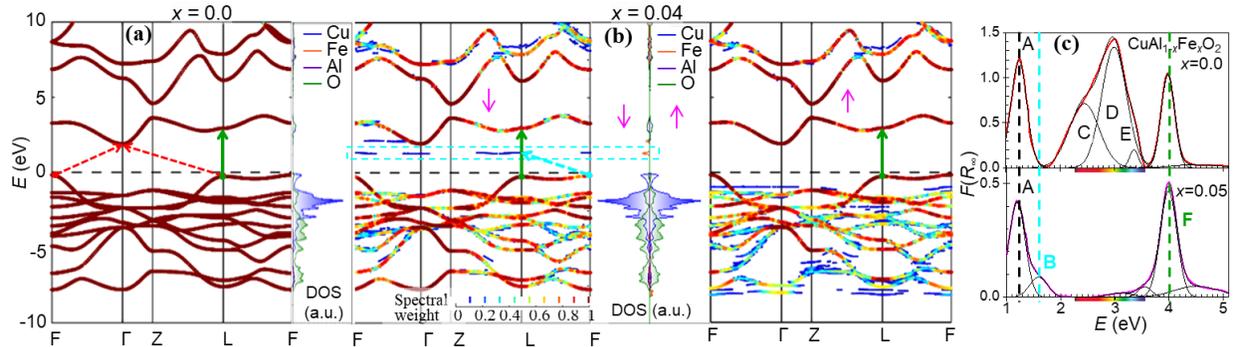

Fig. 2. Electronic band structures of $CuAl_{1-x}Fe_xO_2$, $x = 0.0$ (a) and 0.04 (spin down/up (b)), and $F(R_\infty)$ spectra for $CuAl_{1-x}Fe_xO_2$, $x = 0.0$ and 0.05 (c). **A** at 1.2 eV is associated with defects observed for powders, **C-E** are also defect related. Whereas, **B** at 1.4 eV only in the alloy is associated with the appearance of the Fe- relate conduction band in the spin down and **F** at 4 eV in both spectra is from the direct bandgap.

Phonon dynamics are experimentally measured using Raman and FTIR spectroscopies and compared to the calculated phonon frequencies determined from our weighted dynamical matrix approach [3]. Figure 3.a shows the Raman spectra of the $CuAl_{1-x}Fe_xO_2$ samples ($x = 0$, 0.05, and 0.10). For $x = 0$, there are two prominent lines, identified as $E_g$ at 419 cm$^{-1}$ and $A_{1g}$ at 768 cm$^{-1}$ (from [3]) with $A_{1g}$ involving the vibration of O atoms along the c-axis (parallel to the O–Cu–O bonds) and $E_g$ at 433 cm$^{-1}$, involving the vibration of O atoms perpendicular to the c-axis. The redshift of the $E_g$ and $A_{1g}$ mode for $x$ up to 0.1 are in line with those for $CuAlO_2$ vs. $CuFeO_2$. The main conclusion from these comparisons is that the observed redshifts result from the expansion of the lattice, which in turn enlarges the length of the M–O bonds, reducing their stiffness and hence lowers the frequencies of the Raman modes. FTIR spectroscopy was used to determine the frequencies of IR-active modes of $CuAl_{1-x}Fe_xO_2$. These vibrational modes also follow the predictions made by our model.

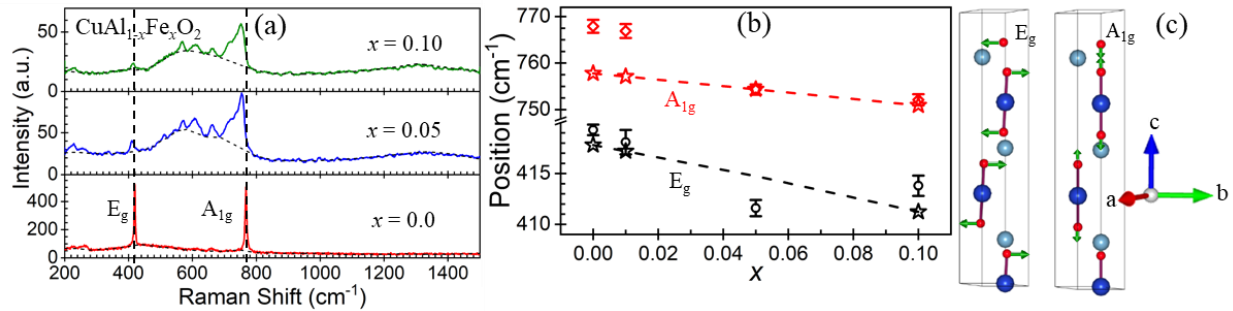

Fig. 3. (a) Raman spectra with $A_{1g}$ and $E_g$ modes. Modes change with Fe-doping due to change in M−O bond. (b) Experimental peak positions of $E_g$ and $A_{1g}$ modes (with error bars) compared to computed results (stars). (c) Schematic views of the $E_g$ and $A_{1g}$.

In this paper, we compare modeled and experimental results for dilute $CuAl_{1-x}Fe_xO_2$ delafossite alloys. The PBEsol+$U$ approach, including Fe spin gives good agreement with the structural and optical properties measured for our synthesized powders. Furthermore, the phonon dynamics experimentally measured using Raman and FTIR spectra agree well with the calculations of our weighted dynamical matrix approach.